\documentclass[12pt]{article}
\usepackage[utf8]{inputenc}
\usepackage{amsmath,bbding,bbm,amsfonts,pdflscape}
\usepackage{setspace}
\usepackage[margin=1in]{geometry}
\newtheorem{assumption}{Assumption}
\newcommand{\ind}{\perp\!\!\!\perp}

\def\transpose{{\sf \scriptscriptstyle{T}}}
\def\trans{^{\transpose}}
\def\inv{^{\sf \scriptscriptstyle{-1}}}

\usepackage{longtable}
\usepackage[tablesonly,notablist]{endfloat}
\usepackage{rotating}
\usepackage{pdflscape}

\DeclareDelayedFloatFlavor{sidewaystable}{table}
\DeclareDelayedFloatFlavour*{longtable}{table}
\DeclareDelayedFloatFlavour*{longsidewaystable}{table}

\usepackage{authblk}
\title{Data Integration in Causal Inference}
\author[1]{Xu Shi}
\author[1]{Ziyang Pan} 
\author[2]{Wang Miao} 
\affil[1]{Department of Biostatistics, University of Michigan, Ann Arbor, MI 48109}
\affil[2]{Department of Probability and Statistics, Peking University, China}
\date{}
\begin{document}

\maketitle
\begin{abstract}
	Integrating data from multiple heterogeneous sources has become increasingly popular to achieve a large sample size and diverse study population. This paper reviews development in causal inference methods that combines multiple datasets collected by potentially different designs from potentially heterogeneous populations. We summarize recent advances on combining randomized clinical trial with external information from observational studies or historical controls, combining samples when no single sample has all relevant variables with application to two-sample Mendelian randomization, distributed data setting under privacy concerns for comparative effectiveness and safety research using real-world data, Bayesian causal inference, and causal discovery methods.
\end{abstract}

\noindent%
{\it Keywords:} Causal inference, data integration, data fusion, generalizability, transportability
\vfill
\newpage
\begin{spacing}{1.5}

\section{Introduction}
The availability of multiple datasets collected by different designs from heterogeneous populations has brought emerging challenges and opportunities for causal inference.
Integrating data from multiple sources to facilitate causal inference has become increasingly popular. 
For example, randomized clinical trial (RCT) has been the gold standard for causal inference but often suffers from insufficient sample size and homogeneous study population due to inclusion/exclusion criteria. 
Results from RCTs may not be generalizable to a real-world population. In contrast, observation study typically offers a diverse sample representative of the target  population with a large sample size but often suffers from unmeasured confounding. 
Combining data from both designs allows one to extend causal inference from an RCT to a target population, to correct for bias in observational studies, and to improve efficiency \cite{colnet2020causal}.
Another prominent example is when no single  dataset contains all relevant variables, that is, there are no complete data for any subject. In this case, identification becomes difficult even for parameters that are straightforward to be identified with complete data \cite{ridder2007econometrics}. 
This is typical in survey sample combination where variables collected in each survey may differ \cite{yang2020statistical}.  This is also the case in two-sample instrumental variable methods, which is widely applied in Mendelian randomization studies where individual-level genetic data are not available due to privacy concerns \cite{angrist1992effect}.

In this paper, we  review  selected literature on data integration methods in causal inference. 
Recent review studies focused on combining randomized and observational data \cite{colnet2020causal,degtiar2021review} and data combination in survey sampling \cite{ridder2007econometrics,yang2020statistical}.  We aim to provide a more systematic review 
and cover a range of research areas.
We start with notation and introduce key assumptions and concepts frequently appeared in the literature in Section~\ref{sec:notation}. We then summarize recent methodological advances in integrating data from   RCTs and observational studies in Section~\ref{sec:trial}, and combining data when no single sample has all relevant variables in Section~\ref{sec:nocompleteddata}. We briefly review the literature on data integration for causal discovery, distributed data analysis for privacy protection, and Bayesian methods for integrated causal inference in Section~\ref{sec:others}. We close with a discussion in Section~\ref{sec:discussion}.

\section{Preliminaries\label{sec:notation}}
In this section, we briefly introduce the potential outcome framework and review key concepts in causal inference and data integration.
Let $A$ denote a binary  treatment $A$ (1: treated, 0: untreated), $Y$ denote an observed outcome, and $X$ denote a vector of measured covariates. When all circumstances are the same except for the treatment status, any difference observed in the outcomes has to be attributed to the treatment. Correspondingly, for each subject we define a pair of potential outcomes,  $(Y(1),Y(0))$, that would be observed if the subject had been given treatment, $Y(1)$, and control, $Y(0)$ \cite{rubin1974estimating}. 
A fundamental problem in causal inference is that for each subject, we can only observe one of the potential outcomes. 
We assume that the treatment is well defined such that the observed outcome is equal to the potential outcome corresponding to the subject's treatment condition, that is
\begin{assumption}[Consistency]
	$Y=Y(a)$ if $A=a$.\label{assump:consistency}
\end{assumption}
Because it is impossible to compute the difference in $Y(1)$ and  $Y(0)$ for a specific subject, we often specify a target population of interest, and study the mean difference in the target population, referred to as the average treatment effect (ATE). 
In practice, we cannot observe data on all subjects in the prespecified target population but rather data on a sample of subjects referred to as the study sample.
Let $S$ be a binary indicator of whether a subject is selected into the study sample (1: sampled, 0: not sampled).  
It is important to note that the ATE is population-specific. 
In fact, we can define multiple ATEs each with respect to a different target population as follows: 
\[
\tau=E[Y(1)-Y(0)], ~~\tau_{\scriptscriptstyle 1} = E[Y(1)-Y(0)\mid S=1], ~~\tau_{\scriptscriptstyle 0} = E[Y(1)-Y(0)\mid S=0].
\] 
For example, the ATE is $\tau$ if the combined $S=1$ and $S=0$ sample is a random sample of the target population. The ATE estimated based on the study sample, i.e., the $S=1$ sample, is an estimate of $\tau_{\scriptscriptstyle 1}$, which is not necessarily equal to $\tau$ because the study sample is not necessarily a representative sample of the target population.
Identification of the ATE, which is a function of the potential outcome distribution in a target population involves expressing it as a function of the observed data distribution, such that distinct data generating mechanisms lead to distinct values.
To identify the ATE, ideally we would like to observe $Y(a)$ of all subjects in the target population to compute $E[Y(a)]$, for $a=0$ or $1$.
However, both sample selection mechanism and treatment assignment mechanism lead to missingness in $Y(a)$: generally $Y(a)$'s are missing for all subjects in the $S=0$ sample (sample selection); with in the $S=1$ sample, $Y(a)$'s are unobserved for subjects in the other treatment arm with $A= a'$, $a'\neq a$ (treatment assignment).
\textit{Confounding bias}, also referred to as violation of \textit{internal validity}, occurs when factors that impact treatment assignment  also predict the outcome, such that the observed $Y(a)$'s in the  $A=a$ arm can not represent the missing $Y(a)$'s in the  $A= a'$ arm, i.e. $E[Y(a)\mid A=a,S=1]\neq E[Y(a)\mid A=a',S=1]$. \textit{Selection bias}, also referred to as violation of \textit{external validity}, occurs when factors that impact sample selection also predict the outcome, such that the observed $Y(a)$'s in the $S=1$ sample cannot represent the missing $Y(a)$'s in the $S=0$ sample. A less stringent condition targeting treatment effect estimation, i.e., the mean difference rather than the mean, defines selection bias as when factors that impact sample selection also modifies the treatment effect, i.e., $E[Y(1)-Y(0)\mid S=1]\neq E[Y(1)-Y(0)\mid S=0]$ \cite{lesko2017generalizing,stuart2011use}.
\textit{Collider-stratification bias} may also occur due to conditioning the analysis on the study sample, if $S$ is a common consequence of the treatment (or a predictor of the treatment) and the outcome (or a predictor of the outcome) \cite{greenland2003quantifying}.

Two key assumptions about the treatment assignment mechanism are often imposed, which we refer to as treatment exchangeability and positivity. The treatment exchangeability assumption states that within a strata of $X$, $Y(a)$ of subjects in the $A=a$ arm can be exchanged with $Y(a)$ of subjects in the $A=a'$  arm:
\begin{assumption}[Treatment exchangeability]
	$Y(a)\ind A\mid X, S=1$.\label{assump:exchangeability}
\end{assumption}
Assumption~\ref{assump:exchangeability} 
allows us to represent the conditional distribution of the unobserved potential outcome using that of the observed potential outcome. 
We thus have that for $a=0$ or $1$,
\begin{equation}
	E[Y(a)\mid X,S=1] \stackrel{Assumption~\ref{assump:exchangeability}}{=} E[Y(a)\mid A=a,X,S=1]\stackrel{Assumption~\ref{assump:consistency}}{=}  E[Y\mid A=a,X,S=1].\label{eq:id1}
\end{equation}
Eq. (\ref{eq:id1}) has also been used as a weaker version of Assumption~\ref{assump:exchangeability}.
Within each strata of the covariates sufficient for the treatment exchangeability, we also need to have nonzero subjects in both treatment arms:
\begin{assumption}[Treatment positivity]\label{assump:positivity}
	$ P(A=a|X, S=1)>0$ for all $a$ almost surely.
\end{assumption}
Often $P(A=a|X, S=1)$ is referred to as the propensity score. Note that Assumptions~\ref{assump:exchangeability} and \ref{assump:positivity} are conditional on the study sample, thus the set of covariates $X$ sufficient for Assumptions~\ref{assump:exchangeability} and \ref{assump:positivity} to hold may include variables beyond common causes of treatment and outcome, i.e., the typical confounders. For example, a covariate that causes selection $S$ and outcome but is independent with the treatment can become a confounder if the treatment also causes selection. This is a consequence of collider-stratification bias where conditioning on $S$ results in a spurious association between the treatment and the covariate.

Besides conditions to ensure internal validity, people often impose another two key assumptions about the sample selection mechanism to ensure external validity, which we refer to as selection exchangeability and positivity, in analogy to  Assumptions~\ref{assump:exchangeability}-\ref{assump:positivity}  \cite{stuart2011use,lesko2017generalizing,dahabreh2020extending}.
\begin{assumption}[Selection exchangeability]
	$Y(a)\ind S\mid X$. \label{assump:exchangeability2}
\end{assumption}
Assumption~\ref{assump:exchangeability2} allows us to generalize the conditional distribution of the potential outcome from the study sample to a target population, such as the one represented by the $S=0$ sample or the combination of $S=0$ and $S=1$ sample:\begin{equation}
	E[Y(a)\mid X,S=1]\stackrel{Assumption~\ref{assump:exchangeability2}}{=} E[Y(a)\mid X,S=0]  \stackrel{Assumption~\ref{assump:exchangeability2}}{=} E[Y(a)\mid X].
	\label{eq:id2}
\end{equation}
Weaker versions of the selection exchangeability assumption include (I) mean conditional exchangeability, i.e. Eq. (\ref{eq:id2}) and (II) all treatment effect modifiers are measured, i.e., $E[Y(1)-Y(0)\mid X]=E[Y(1)-Y(0)\mid X,S=1]$.
We further assume that variables required for selection exchangeability do not serve as study eligibility criteria that completely exclude certain subjects from the study sample. 
\begin{assumption}[Selection positivity]\label{assump:positivity2}
	$P(S=s|X)>0$ for all $s$ almost surely.
\end{assumption}
For example, suppose geographic location restricted study participation such that there is zero probability of selecting subjects in certain area, then Assumption~\ref{assump:positivity2} requires that geographic location is not needed for Assumption~\ref{assump:exchangeability2}, i.e., conditional on $X$, geographic location is not associated with the outcome or does not modify the treatment effect.

\section{Combining randomized clinical trial with external data\label{sec:trial}}
There is a rich literature on combining information from both experimental and non-experimental designs and bridging findings from an RCT to a target population \cite{cole2010generalizing,o2014generalizing,tipton2013improving,hartman2015sate,lesko2017generalizing,rudolph2017robust,westreich2017transportability,buchanan2018generalizing,dahabreh2019extending,dahabreh2019study,dahabreh2019generalizingtrial,dahabreh2017Biometricsgeneralizing,dahabreh2019generalizing,dahabreh2019efficient,dahabreh2020toward,dahabreh2020extending,dong2020integrative}. 
In this setting, $S=1$ indicates the sample of trial participants, and we observe $(Y,A,X,S=1)$ in RCT data. Due to randomization or stratified randomization, the propensity score, $P(A=a\mid X,S=1)$ is a known function designed by the investigator, and Assumptions~\ref{assump:exchangeability}-\ref{assump:positivity} naturally hold with $X$ being the variables defining the strata. 

Two problems are frequently studied: 
generalizability \cite{cole2010generalizing,stuart2011use,dahabreh2017Biometricsgeneralizing,buchanan2018generalizing} and transportability \cite{pearl2014external,bareinboim2016causal,westreich2017transportability,rudolph2017robust,hunermund2019causal}. 
The distinction between the two concepts is well summarized in \cite{dahabreh2019extending} and \cite{degtiar2021review}: \textit{generalizability} focuses on the setting when the study sample is a subset of the target population 
, and \textit{transportability} considers the setting when the study sample and the target population are partially- or non-overlapping. 
An example of the generalizability problem is: suppose the target population is the trial-eligible population, and the combined $S=1$ and $S=0$ sample is a random sample of the target population, in which trial participants are in the $S=1$ sample and non-participants are in the $S=0$ sample. In this case, the target ATE is $\tau$ and we would like to generalize  inference about $\tau_{\scriptscriptstyle 1}$ obtained from the trial data to $\tau$. 
An example of the transportability problem is: suppose the target population is a real-world population, and  $S=0$ sample is a random sample of the target population separately obtained from external data sources such as administrative healthcare  databases or survey studies. In this case, the target ATE is $\tau_{\scriptscriptstyle 0}$ and we would like to transport inference about $\tau_{\scriptscriptstyle 1}$ to $\tau_{\scriptscriptstyle 0}$. 

Both  problems require some information in the $S=0$ sample, and often two scenarios are considered: (S1) covariates are measured on all individuals in the $S=0$ sample, i.e., we observe $(X,S=0)$; (S2) covariates are measured on a subsample of the $S=0$ sample, i.e., we observe $(X,S=0,D=1)$, where $D$ indicates whether we have data on $X$. 
In scenario (S2), it is often assumed that $D\ind (Y,A,X)\mid S$ such that $P(D=1\mid Y,A,X,S)=P(D=1\mid S)$. That is, $(X,S=0,D=1)$ is a simple random sample of the $S=0$ sample with two possibilities: (S2.1) $P(D=1\mid S=0)$ is a known constant; (S2.2) $P(D=1\mid S=0)$ is an unknown constant. \cite{dahabreh2019study} and \cite{dahabreh2019generalizing} showed that $\tau$ is not identifiable under (S2.2), while $\tau_{\scriptscriptstyle 0}$ is always identifiable in (S2). 

\subsection{Generalizability and transportability methods \label{sec:general}}
In this section, we review three common strategies for identification and estimation of $E[Y(a)]$ (generalizability) and $E[Y(a)\mid S=0]$ (transportability) for $a=0$ or $1$. Correspondingly, the ATE $\tau$ and $\tau_{\scriptscriptstyle 0}$ can be directly obtained based on $E[Y(a)]$ and $E[Y(a)\mid S=0]$ by definition.
To illustrate the methods, we take scenario (S1) as an example where 
we observe $(Y_i,A_i,X_i,S_i=1)_{\scriptscriptstyle i=1}^{\scriptscriptstyle n_{\scriptscriptstyle 1}}$ and $(X_i,S_i=0)_{\scriptscriptstyle i=n_{\scriptscriptstyle 1}+1}^{\scriptscriptstyle n_{\scriptscriptstyle 1}+n_{\scriptscriptstyle 0}}$ from a total of $n=n_{\scriptscriptstyle 1}+n_{\scriptscriptstyle 0}$ subjects. We summarize the methods under all scenarios in Table~\ref{tbl:RctObs}.

	\begin{sidewaystable}
			\begin{tabular}{|c|c|c|}
				\hline 
				& \textbf{Generalizability ($E[Y(a)]$)} & \textbf{Transportability ($E[Y(a)\mid S=0]$)}   \tabularnewline
				\hline 
				\multicolumn{3}{|c|}{\textbf{(S1) Covariates are measured on all individuals in the $S=0$ sample,
						i.e., we have $(X,S=0)$}}\tabularnewline
				\hline 
				\textbf{OR} & $E\{m_a(X)\}$ & $E\{m_a(X)\mid S=0\}=E\{\frac{\mathbb{I}(S=0)}{P(S=0)}m_a(X)\}$\tabularnewline
				\hline 
				\textbf{IPW} & $E\{\frac{\mathbb{I}(S=1,A=a)}{P(S=1,A=a\mid X)}Y\}$ 
				\footnotemark[1] & $E\{\frac{\mathbb{I}(S=1,A=a)P(S=0\mid X)}{P(S=1,A=a\mid X)P(S=0)}Y\}=\frac{E\{\frac{\mathbb{I}(S=1,A=a)P(S=0\mid X)}{P(S=1,A=a\mid X)}Y\}}{E\{\frac{\mathbb{I}(S=1,A=a)P(S=0\mid X)}{P(S=1,A=a\mid X)}\}}$
				\footnotemark[2]\tabularnewline
				\hline 
				\textbf{AIPW} & $E\{\frac{\mathbb{I}(S=1,A=a)}{P(S=1,A=a\mid X)}(Y-m_a(X))+m_a(X)\}$ & $E\{\frac{\mathbb{I}(S=1,A=a)P(S=0\mid X)}{P(S=1,A=a\mid X)P(S=0)}(Y-m_a(X))+\frac{\mathbb{I}(S=0)}{P(S=0)}m_a(X)\}$\tabularnewline
				\hline 
				\multicolumn{3}{|c|}{\textbf{(S2.1) Covariates are measured on all individuals in the $S=0$ sample and $P(D=1\mid S=0)$ is known
				}}\tabularnewline
				\hline 
				\textbf{OR} & $E[E\{m_a(X)\mid S,D=1\}]=E\{\frac{\mathbb{I}(D=1)}{P(D=1\mid S)}m_a(X)\}$
				\footnotemark[3] & $E\{m_a(X)\mid S=0,D=1\}$ 
				\footnotemark[4]\tabularnewline
				\hline 
				\textbf{IPW} & $E\{\frac{\mathbb{I}(S=1,A=a)}{P(S=1,A=a\mid X)}Y\}$ 
				\footnotemark[5] & $E\{\frac{\mathbb{I}(S=1,A=a)P(S=0\mid X)}{P(S=1,A=a\mid X)P(S=0)}Y\}$
				\footnotemark[6]\tabularnewline
				\hline 
				\textbf{AIPW} & $E\{\frac{\mathbb{I}(S=1,A=a)}{P(S=1,A=a\mid X)}(Y-m_a(X))+\frac{\mathbb{I}(D=1)}{P(D=1\mid S)}m_a(X)\}$ & $E\{\frac{\mathbb{I}(S=1,A=a)P(S=0\mid X)}{P(S=1,A=a\mid X)P(S=0)}(Y-m_a(X))+\frac{\mathbb{I}(S=0,D=1)}{P(S=0,D=1)}m_a(X)\}$\tabularnewline
				\hline 
				\multicolumn{3}{|c|}{\textbf{(S2.2) Covariates are measured on a subsample of the $S=0$ sample and $P(D=1\mid S=0)$ is unknown
				}}\tabularnewline
				\hline 
				\textbf{OR} & Not identifiable 
				\footnotemark[7] & $E\{m_a(X)\mid S=0,D=1\}$\tabularnewline
				\hline 
				\textbf{IPW} & Not identifiable 
				\footnotemark[8] & 
				$E\{\frac{\mathbb{I}(S=1,A=a)P(S=0\mid X,D=1)}{P(A=a\mid S=1,X)P(S=1\mid X,D=1)P(S=0,D=1)}Y\}$
				\footnotemark[9]\tabularnewline
				\hline 
				\textbf{AIPW} & Not identifiable &
				$E\{\frac{\mathbb{I}(S=1,A=a)P(S=0\mid X,D=1)}{P(A=a\mid S=1,X)P(S=1\mid X,D=1)P(S=0,D=1)}(Y-m_a(X))+\frac{\mathbb{I}(S=0,D=1)}{P(S=0,D=1)}m_a(X)\}$
				\tabularnewline
				\hline 
			\end{tabular}
			\footnotetext{\footnotesize \hspace{-0.25in}(1) $P(S=1,A=a\mid X)=P(A=s\mid S=1,X)P(S=1\mid X)$ where $P(A=s\mid S=1,X)$ is designed by the investigator in an RCT and can also be estimated based on the $S=1$ sample, and $P(S=1\mid X)$ identified
				from the combined sample.\\
				(2) $E\{\frac{\mathbb{I}(S=1,A=a)P(S=0\mid X)}{P(S=1,A=a\mid X)}\}=P(S=0)$,
				hence $E\{\frac{\mathbb{I}(S=1,A=a)P(S=0\mid X)}{P(S=1,A=a\mid X)P(S=0)}Y\}=E\{\frac{\mathbb{I}(S=1,A=a)P(S=0\mid X)}{P(S=1,A=a\mid X)}Y\}/E\{\frac{\mathbb{I}(S=1,A=a)P(S=0\mid X)}{P(S=1,A=a\mid X)}\}$.\\
				(3) $P(D=1\mid S)=\mathbb{I}(S=1)+\mathbb{I}(S=0)P(D=1\mid S=0)$.\\
				(4) $E\{m_a(X)\mid S=0\}=E\{m_a(X)\mid S=0,D=1\}$ because $D\perp X\mid S$.\\
				(5) $P(S=1\mid X)$ identified by $\frac{P(S=1\mid X)}{P(S=0\mid X)}=\frac{P(S=1\mid X,D=1)}{P(S=0\mid X,D=1)}P(D=1\mid S=0)$. Estimation strategies are proposed in \cite{dahabreh2019study}. \\
				(6) Similar to footnote (5), 
				$P(S=1)$ identified by $\frac{P(S=1)}{P(S=0)}=\frac{P(S=1\mid D=1)}{P(S=0\mid D=1)}P(D=1\mid S=0)$.\\
				(7) Unlike footnote (3), $P(D=1\mid S)$
				is not identifiable because $P(D=1\mid S=0)$ is unknown.\\
				(8) Unlike footnote (5), $P(S=1\mid X)$ is not identifiable
				because $P(D=1\mid S=0)$ is unknown.\\
				(9) By footnote (1) and (5), $
				E\{\frac{\mathbb{I}(S=1,A=a)P(S=0\mid X)}{P(S=1,A=a\mid X)P(S=0)}Y\}
				=
				E\{\frac{\mathbb{I}(S=1,A=a)P(S=0\mid X,D=1)}{P(A=a\mid S=1,X)P(S=1\mid X,D=1)P(S=0,D=1)}Y\}$. 
				Although $P(S=1\mid X)$ is not identifiable as shown in footnote (8), $P(S=1\mid X,D=1)$ is identifiable.  
			} 
			\caption{\label{tbl:RctObs}Three estimation strategies (OR = outcome regression, IPW = inverse probability weighting, AIPW = augmented IPW) for the ATE in the target population, when either the combined sample (generalizability, the ATE is $\tau$) or the $S=0$ sample  (transportability, the ATE is $\tau_{\scriptscriptstyle 0}$) is a random sample of the target population. We present three representation of $E[Y(a)]$ and $E[Y(a)\mid S=0]$, $a=0$ or $1$, to simplify exposition.}
	\end{sidewaystable}

\subsubsection{Outcome regression (OR)\label{sec:or}}
Let $m_{\scriptscriptstyle a}(x)=E[Y\mid A=a,X=x, S=1]$ denote the conditional mean outcome in the study sample and $\widehat{m}_a(x)$ denote an estimated model using  $(Y_i,A_i,X_i,S_i=1)_{\scriptscriptstyle i=1}^{\scriptscriptstyle n_{\scriptscriptstyle 1}}$.
Under Assumptions~\ref{assump:consistency}-\ref{assump:positivity2}, we have the following identification result
\begin{equation}
	\begin{split}
		&E[Y(a)] = E[m_a(X)] =\int_{\mathcal{X}}m_a(x)f(x)dx,~~\text{ and }~~\\
		&E[Y(a)\mid S=0] = E[m_a(X)\mid S=0] =\int_{\mathcal{X}}m_a(x)f(x\mid S=0)dx.
		\label{eq:id_gform}
	\end{split}
\end{equation}
Both $f(x)$ and $f(x\mid S=0)$ are identifiable in scenario (S1) where we have observed $X$ on all individuals. 
Therefore, we can marginalize $\widehat{m}_a(x)$ over the empirical distribution of $X$ in the combined sample and the $S=0$ sample, respectively, which gives the following outcome regression estimators \cite{lesko2017generalizing,dahabreh2017Biometricsgeneralizing,dahabreh2019generalizing}
\begin{equation}
	\begin{split}
		&\widehat{E}[Y(a)] = \frac{1}{n}\sum_{i=1}^n\widehat{m}_a(X_i),~~\text{ and }~~\\
		&\widehat{E}[Y(a)\mid S=0] = \frac{1}{n_{\scriptscriptstyle 0}} \sum_{i=1}^{n_{\scriptscriptstyle 0}}\widehat{m}_a(X_i) =
		\frac{1}{n}\sum_{i=1}^n
		\frac{\mathbb{I}(S_i=0)}{\widehat{P}(S=0)}\widehat{m}_a(X_i),
	\end{split}
\end{equation}
where $\widehat{P}(S=0)=n^{\scriptscriptstyle -1}\sum_{i=1}^n\mathbb{I}(S_i=0)=n_{\scriptscriptstyle 0}/n$. 
Eq. (\ref{eq:id_gform}) has been referred to as the g-formula \cite{greenland1986identifiability,robins1986new} or  standardization \cite{vansteelandt2011invited}  in epidemiology, and can also be viewed as  imputation in missing data literature \cite{full_imputation}.

\subsubsection{Inverse probability weighting (IPW)\label{sec:ipw}}
Inverse probability weighting is a very commonly used technique \cite{cole2010generalizing,lesko2017generalizing,westreich2017transportability,dahabreh2017Biometricsgeneralizing,dahabreh2019generalizing}. Note that the g-formula in Eq. (\ref{eq:id_gform}) can be re-expressed as follows
\begin{equation}
	\begin{split}
		&E[m_a(X)] =E[\frac{\mathbb{I}(S=1,A=a)}{P(S=1,A=a\mid X)}Y],~~\text{ and }~~\\
		&E[m_a(X)\mid S=0] =E[\frac{\mathbb{I}(S=1,A=a)P(S=0\mid X)}{P(S=1,A=a\mid X)P(S=0)}Y],
		\label{eq:id_ipw}
	\end{split}
\end{equation}
where $P(S=1,A=a\mid X)=P(A=a\mid S=1,X)P(S=1\mid X)$. The propensity score, $P(A=a\mid S=1,X)$, is a known function designed by the investigator in an RCT, while the trial participation probability $P(S=1\mid X)$ can be estimated in the combined sample because $X$ is fully observed under (S1).
We arrive at the following inverse probability weighting estimators
\begin{equation}\label{eq:IPW}
	\begin{split}
		&\widehat{E}[Y(a)] = \frac{1}{n}\sum_{i=1}^n \frac{\mathbb{I}(S_i=1,A_i=a)Y_i}{\widehat{P}(A=a, S=1\mid X_i)},~~\text{ and }~~\\
		&\widehat{E}[Y(a)\mid S=0] = 
		\frac{1}{n}\sum_{i=1}^n
		\frac{\mathbb{I}(S_i=1,A_i=a)[1-\widehat{P}(S=1\mid X_i)]Y_i}{\widehat{P}(A=a, S=1\mid X_i)\widehat{P}(S=0)},
	\end{split}
\end{equation}
where $\widehat{P}(S=0)=n^{\scriptscriptstyle -1}\sum_{i=1}^n\mathbb{I}(S_i=0)$ and $\widehat{P}(A=a, S=1\mid X)=\widehat{P}(A=a\mid S=1,X)\widehat{P}(S=1\mid X)$ is a product of the estimated treatment and trial participation probabilities. Although the propensity score is known, estimating the model parameters rather than using the true value can improve efficiency \cite{hahn1998role,lunceford2004stratification}.
Comparing Eq. (\ref{eq:IPW}) to traditional IPW estimator using the trial data only, i.e., 
\begin{equation}
	\widehat{E}[Y(a)\mid S=1] = \frac{1}{n}\sum_{i=1}^n \frac{\mathbb{I}(S_i=1,A_i=a)Y_i}{\widehat{P}(A=a\mid S=1, X_i)},
\end{equation}
we further weight each subject who participated in the trial by the inverse of the trial participation probability, $P(S=1\mid X)$, to generalize the ATE from the $S=1$ sample to the combined sample, while to transport the ATE from the $S=1$ sample to the $S=0$ sample, trail participants are weighted by the inverse of both the odds of trial participation $P(S=1\mid X)/P(S=0\mid X)$ and $P(S=0)$.

\subsubsection{Augmented inverse probability weighting (AIPW)}
So far, each of the estimators relies on estimating components of the likelihood such as $m_a(X)$ and $P(S=1,A=a\mid X)$, which are  not necessarily in themselves of scientistic interest. Nonparametric estimation may not be feasible when $X$ is of high dimension, while dimension reducing working models may be prone to model misspecification.  We can combine the two estimators to gain robustness. A common approach to derive a robust estimator is by constructing an estimating equation from the efficient influence function (EIF) and  evaluating it under a working model for the observed data distribution to solve for the parameter of interest, which is widely used in missing data problems \cite{tsiatis2007semiparametric}. 
Any regular and asymptotic linear estimator is asymptotically equivalent to the sample average of the influence function, which is a function of the observed data with mean zero and finite variance, and the one with the smallest variance is referred to as the EIF \cite{van2000asymptotic,tsiatis2007semiparametric}.
The EIFs for $E[Y(a)]$ and $E[Y(a)\mid S=0]$ are
\begin{equation}
	\begin{split}
		U(O; E[Y(a)])&=\frac{\mathbb{I}(S=1,A=a)}{P(S=1,A=a\mid X)}(Y-m_a(X))+m_a(X)-E[Y(a)],\\
		U_{\scriptscriptstyle 0}(O; E[Y(a)\mid S=0])&= \frac{\mathbb{I}(S=1,A=a)P(S=0\mid X)}{P(S=1,A=a\mid X)P(S=0)}(Y-m_a(X))\\
		&+\frac{\mathbb{I}(S=0)}{P(S=0)}m_a(X) - E[Y(a)\mid S=0],
	\end{split}
	\label{eq:id_aipw}
\end{equation}
where $O=(S\times Y,S\times A,X,S)$ denotes the observed data, and $E[U(Y,A,X,S; E[Y(a)])]=E[U_{\scriptscriptstyle 0}(Y,A,X,S; E[Y(a)\mid S=0])]=0$ at the true values. Let $\widehat{U}(\cdot)$ and $\widehat{U}_{\scriptscriptstyle 0}(\cdot)$ respectively denote the evaluation of $U(\cdot)$ and ${U}_{\scriptscriptstyle 0}(\cdot)$ under an estimated working model, then we can obtain the AIPW estimators by solving $\sum_{i=1}^n \widehat{U}(O_i; E[Y(a)])/n =0$ and $\sum_{i=1}^n \widehat{U}_{\scriptscriptstyle 0}(O_i; E[Y(a)\mid S=0])/n=0$ \cite{dahabreh2017Biometricsgeneralizing,dahabreh2019generalizing}. 
As mentioned in Section~\ref{sec:ipw}, $P(A=a\mid S=1,X)$ is guaranteed to be correctly specified in an RCT, therefore $P(A=a, S=1\mid X)$ is correctly specified as long as $P(S=1\mid X)$ is.
Hence the above AIPW estimators are doubly robust in the sense that it remains consistent when either the probability of trial participation $P(S=1\mid X)$ or the outcome regression model $m_a(X)$ is correctly specified.
This can be seen by the following observation:  the IPW estimator introduced in Section~\ref{sec:ipw} can be obtained by misspecifying $m_a(X)$ as zero in Eq. (\ref{eq:id_aipw}), while the OR estimator introduced in Section~\ref{sec:or} can be obtained by setting the weight in the first term of both $U(\cdot)$ and ${U}_{\scriptscriptstyle 0}(\cdot)$ to zero in Eq. (\ref{eq:id_aipw}).

\subsubsection{Other methods for combining data from clinical trial and external data}
Other doubly robust estimators include a targeted maximum likelihood estimator \cite{rudolph2017robust} and a  augmented calibration weighting estimator \cite{dong2020integrative}.
Sensitivity analysis that replaces Assumption~\ref{assump:exchangeability2} with a pre-specified bias function has also been proposed \cite{dahabreh2019sens}. 
Meta-analysis is often used to synthesize information about parameters from data collected from multiple trials, which allows for extensions of the above methods to the setting of generalizing or transporting inferences from multiple randomized RCTs to a target population  \cite{dahabreh2019efficient,manski2000identification,steele2020importance,dahabreh2020toward}. Identification under an arbitrary collection of observational and experimental data 
has been investigated \cite{lee2020general}. Combining probability and non-probability samples with high dimensional data has also been studied \cite{yang2020doubly}.

\subsection{Correcting for bias in observational study using validation or trial data\label{sec:correctobs}}
Internal validity, i.e., Assumptions~\ref{assump:exchangeability}-\ref{assump:positivity}, naturally hold in RCTs due to randomization but not necessarily in observational studies due to potential  unmeasured confounding. 
Borrowing strength from the internal validity of RCT data and the large sample size of observation data can mitigate bias and improve efficiency. 

In this vein, \cite{yang2020improved} considered estimation of the average treatment effect on the treated (ATT) in the scenario where $X=(X_{\scriptscriptstyle 1},U)$, and $U$ is unobserved. Data are obtained from RCT $(Y, A,X_{\scriptscriptstyle 1}, S=1)$ and from observational study $(Y,A,X_{\scriptscriptstyle 1},S=0)$. In RCT, $X_{\scriptscriptstyle 1}$ is sufficient for Assumption~\ref{assump:exchangeability}, while in the observational study, the unmeasured confounding $U$ leads to bias. A weaker version of Assumption~\ref{assump:exchangeability2} is further assumed.
\cite{yang2020improved} proposed to model unmeasured confounding bias via $\lambda(X_{\scriptscriptstyle 1};\phi)=E[Y(0)\mid A=1,X_{\scriptscriptstyle 1},S=0;\phi]-E[Y(0)\mid A=0,X_{\scriptscriptstyle 1},S=0;\phi]$, which is equal to zero if $U=\emptyset$. 
Modeling this bias function allows one to improve efficiency in estimation of the ATT by combining observational data and RCT data. A similar idea was considered in \cite{kallus2018removing} where a confounding bias correction term was learned with interpolation of $E[Y\mid A, X_{\scriptscriptstyle 1}]$ between RCT and observational data, and \cite{gui2020combining} where RCT data were used to correct bias in an imperfect estimator based on an invalid instrumental variable defined on observation data.

In \cite{athey2020combining}, it is assumed that we observe data from RCT $(W, A, X, S=1)$ and from observational study $(Y,W, A,X, S=0)$, where $W$ denotes a secondary outcome observed in both studies, $Y$ denotes the primary outcome expensive to measure in RCT, and the $S=0$ sample is a random sample of the target  population. 
Motivated by the observation that the treatment effects on the secondary outcome should be similar in the RCT and observational data if $X$ is sufficient for Assumption~\ref{assump:exchangeability}, \cite{athey2020combining} developed a control function method for using differences in the estimated causal effects on the secondary outcome between the two samples to adjust estimation of the treatment effect on the primary outcome. 

\cite{yang2019combining} considered the scenario where a small validation dataset with all confounders $(Y, A,X_{\scriptscriptstyle 1},U, S=1)$ and a big main dataset with unmeasured confounders $(Y,A,X_{\scriptscriptstyle 1},S=0)$ are available. Both are random samples of the target population hence external validity is satisfied. The big main data can improve efficiency and the small validation data can ensure consistency.
For each dataset $S=s$, let $\widehat{\tau}_{\scriptscriptstyle \text{s}},~s=0,1$ denote a consistent estimator of the ATE based on a user-specified estimation strategy adjusting for all confounders $(X_{\scriptscriptstyle 1},U)$, and let  $\widehat{\tau}_{\scriptscriptstyle \text{ep, s}},~s=0,1$ denote an error-prone estimator using the same estimation strategy but with $U$ uncontrolled. Apparently $\widehat{\tau}_{\scriptscriptstyle \text{0}}$ can not be obtained.
A key insight is that the two error-prone estimates $\widehat{\tau}_{\scriptscriptstyle \text{ep, 1}}-\widehat{\tau}_{\scriptscriptstyle \text{ep, 0}}$ should be consistent for zero. By modeling the joint distribution of $\widehat{\tau}_{\scriptscriptstyle \text{1}}$ and $\widehat{\tau}_{\scriptscriptstyle \text{ep, 1}}-\widehat{\tau}_{\scriptscriptstyle \text{ep, 0}}$, they derived the most efficient consistent estimator of $\tau$ among all linear combinations $\widehat{\tau}_{\scriptscriptstyle \text{1}}+\alpha(\widehat{\tau}_{\scriptscriptstyle \text{ep, 1}}-\widehat{\tau}_{\scriptscriptstyle \text{ep, 0}}), \alpha\in R$.  Other methods for controlling unmeasured confounding with validation data include the propensity score calibration \cite{sturmer2005adjusting} and conditional propensity scores \cite{mccandless2012adjustment}.

\subsection{Combining clinical trial with external control}\label{sec:external}
Single-arm clinical trials are typically conducted for rare diseases due to difficulties in recruiting enough patients for an adequately powered two-arm trial, or for diseases with high unmet medical need that raise ethical concerns  \cite{cuffe2011inclusion,viele2014use,abrahami2021use}. 
Historical or contemporaneous information on the control arm are often available from previous RCT or observational studies. Such external controls have been used to emulate the control arm in the setting of single-arm trials, which can decrease costs and duration and improve power. 

Formally, the single-arm trial data  $(Y,A=1,X,S=0)$ are a random sample of the target population, while the external control data contain $(Y,A=0,X,S=1)$. Our goal is to estimate $E[Y(0)\mid S=0]$ leveraging historical data in order to contrast it with the mean response in the single-arm trial to estimate the treatment effect.
Traditional methods to account for differences in patient characteristics between the external control and the target population 
include meta-analysis \cite{schmidli2014robust,hasegawa2017myth,weber2018use,zhang2019bayesian,schmidli2020beyond} and matching \cite{signorovitch2010comparative,schmidli2020beyond}. Typically a form of exchangeability across different studies like Assumption~\ref{assump:exchangeability2} is assumed. 
Recently, \cite{li2020target} proposed to build an outcome regression model using external control data under exchangeability, and then estimate $E[Y(0)\mid S=0]$ by standardization, which is similar to the identification strategy in Eq. (\ref{eq:id_gform}) with $a=0$.
Besides single-arm trial data, external controls have also been used to improve efficiency in a traditional RCT with data on both arms available. \cite{li2020improving} showed that the semiparametric efficiency bound for estimating $E[Y(1)-Y(0)\mid S=0]$ is reduced by incorporating external control data, and  proposed  a doubly robust and locally efficient estimator that combines outcome regression and inverse probability of treatment weighting.

\section{No single sample contains all relevant variables \label{sec:nocompleteddata}}
\def\zz{Z}
\def\xx{X}
\def\yy{Y}
The data integration problems described so far have complete data on all relevant variables in at least one sample. 
A more challenging problem is when there are no  complete data at any data source. 
This setting has been referred to as data combination \cite{ridder2007econometrics,shu2020improved} or data fusion \cite{evans2018doubly,sun2018semiparametric,li2020causal} in the literature. In the following, we will first introduce methods applicable to the general data combination problem in Section~\ref{sec:datafusion}. We will use a new set of notation in Section~\ref{sec:datafusion} while notation in the rest of the paper follows Section~\ref{sec:notation}. We will then overview specific causal inference problems and methods in Sections~\ref{sec:tsiv}-\ref{sec:datafusion_causal}.

\subsection{General data combination methods\label{sec:datafusion}}%
We first introduce some new notation. Suppose for each member from a population of interest, we can define a vector of relevant variables $(\yy,\xx,\zz)$. 
A sample of complete data on $(\yy,\xx,\zz)$ is unavailable, instead two separate samples are available.
In one sample we observe variables $(\zz,\yy,S=1)$ and in the other sample we observe  $(\zz,\xx,S=0)$, with $\zz$ shared by the two datasets. Suppose the $S=1$ and $S=0$ samples are of size $n_{\scriptscriptstyle 1}$ and $n_{\scriptscriptstyle 0}$, respectively, with total sample size $n=n_{\scriptscriptstyle 1}+n_{\scriptscriptstyle 0}$, then a merged sample combining the two samples is an i.i.d. sample containing $(\yy_iS_i,\zz_i,\xx_i(1-S_i),S_i)_{i=1}^{n}$.

\subsubsection{Estimation of general parameters defined through moment restrictions \label{sec:moment}}
We assume that the $S=1$ sample is drawn from the population of interest, while the $S=0$ sample is an auxiliary sample independent of the $S=1$ sample, which ensures identification that could not be achieved by the $S=1$ sample alone.
We are often interested in a population parameter defined as the unique solution $\theta\in\mathcal{R}^k$ to the $k\times 1$ vector of population moment conditions $E[m(\yy,\xx,\zz;\theta)\mid S=1]=0$, which includes the maximum likelihood estimation and generalized method of moments as special cases. 
For example, $\theta$ is the ATT when $S$ is the binary treatment indicator, $(\yy,\xx)$ are the potential outcomes under treatment and control respectively,  $\zz$ is a vector of pretreatment covariates, and $m(\yy,\xx,\zz;\theta)=\yy-\xx-\theta$. 
Another example is the two-sample instrumental variable (IV) problem, where $\zz$ is a vector of IVs, $\xx$ is the treatment (not necessarily binary), $\yy$ is the outcome, and $m(\yy,\xx,\zz;\theta=(\theta_{\scriptscriptstyle 0},\theta_{\scriptscriptstyle 1}))=\zz(\yy-\theta_{\scriptscriptstyle 0}\xx-\theta_{\scriptscriptstyle 1}\trans \zz)$. We will detail the two-sample IV literature in Section~\ref{sec:tsiv}. Typically selection exchangeability ($S\ind (\yy,\xx)\mid \zz$) and positivity ($P(S=s\mid \zz)>0$) are assumed to identify $\theta$ combining the two samples.

\cite{graham2016efficient} and \cite{shu2020improved} proposed doubly robust and locally efficient estimators of $\theta$ extending the semiparametric efficiency theory of \cite{hahn1998role} and \cite{chen2008semiparametric}. We illustrate the estimation strategies in \cite{shu2020improved} below. When $\yy=\emptyset$, the moment restriction becomes $E[m(\xx,\zz;\theta)\mid S=1]=0$ in which $X$ is unobserved in the $S=1$ sample and we need to combine the two samples for estimation.
\cite{shu2020improved} took the EIF in \cite{chen2008semiparametric} as the estimating function to obtain an AIPW estimator, which solves $\sum_{i=1}^{n} \widehat{U}(S_i,\xx_i,\zz_i;m(\cdot),\theta)/n=0$ where 
{
	\small{
	\begin{equation}
		U(S,\xx,\zz;m(\cdot),\theta)=
		SE[m(\xx,\zz;\theta)\mid \zz]+
		\frac{(1-S)P(S=1\mid \zz)}{P(S=0\mid \zz)}\{m(\xx,\zz;\theta)-E[m(\xx,\zz;\theta)\mid \zz]\}
		.\label{eq:shu}
\end{equation}
}
}
The AIPW estimator is doubly robust in that it remains consistent when either the propensity score model $P(S=1\mid \zz)$ or the outcome regression model $E[m(\xx,\zz;\theta)\mid \zz]$ is correctly specified.
This can be seen by the following observation:  an IPW estimator can be obtained by misspecifying $E[m(\xx,\zz;\theta)\mid \zz]$ as zero in Eq. (\ref{eq:shu}), while an outcome regression estimator can be obtained by setting $P(S=1\mid \zz)/P(S=0\mid \zz)$ to zero in Eq. (\ref{eq:shu}).

When $\yy\neq \emptyset$, \cite{graham2016efficient} and \cite{shu2020improved} further imposed a key identification assumption that the moment condition is separable in the sense that
$E[m(\yy,\xx,\zz;\theta)\mid S=1]=E[m_{\scriptscriptstyle 1}(\yy,\zz;\theta)-m_{\scriptscriptstyle 0}(\xx,\zz;\theta)\mid S=1]$, 
where $m_{\scriptscriptstyle 1}$ and $m_{\scriptscriptstyle 0}$ only depend on variables observed in one sample. We can see that $E[m_{\scriptscriptstyle 1}(\yy,\zz;\theta)\mid S=1]$ can be directly estimated from the $S=1$ sample, while the challenge is to estimate $E[m_{\scriptscriptstyle 0}(\xx,\zz;\theta)\mid S=1]$ combining both samples. Motivated by the observation that estimation of $E[m_{\scriptscriptstyle 0}(\xx,\zz;\theta)\mid S=1]$ reduces to the $\yy=\emptyset$ case with $m(\cdot)$ substituted with $m_{\scriptscriptstyle 0}(\cdot)$,
\cite{shu2020improved} proposed an AIPW estimator that solves $\sum_{i=1}^{n} \widehat{U}_{\scriptscriptstyle 1}(S_i,\yy_i,\xx_i,\zz_i;m_{\scriptscriptstyle 1}(\cdot),m_{\scriptscriptstyle 0}(\cdot),\theta)/n=0$ where 
\begin{equation*}
	{U}_{\scriptscriptstyle 1}(S,\yy,\xx,\zz;m_{\scriptscriptstyle 1}(\cdot),m_{\scriptscriptstyle 0}(\cdot),\theta)=Sm_{\scriptscriptstyle 1}(\yy,\zz;\theta) - U(S,\xx,\zz;m_{\scriptscriptstyle 0}(\cdot),\theta),
\end{equation*}
with $U(S,\xx,\zz;m_{\scriptscriptstyle 0}(\cdot),\theta)$ being the estimating function in Eq. (\ref{eq:shu}) with $m(\cdot)$ substituted with $m_{\scriptscriptstyle 0}(\cdot)$.

An alternative  assumption often imposed is the conditional independence assumption, i.e., $\yy\ind \xx \mid \zz$ \cite{ridder2007econometrics,ogburn2020awarning}. Under this assumption we have $f(\yy,\xx,\zz)=f(\yy\mid \zz)f(\xx,\zz)=f(\xx\mid \zz)f(\yy,\zz)$ where each of $f(\yy, \zz)$ and $f(\xx,\zz)$  can  be estimated from one sample. Therefore, the sample moment conditions can be computed combining the two samples.

\subsubsection{Statistical matching}
Another set of methods in data combination problems is statistical matching, which has been proposed mainly under two scenarios. In the first scenario, a sufficient number of units are shared between the two data sources, i.e., the two samples are partially overlapping. In this case, it is convenient to merge the two samples by linking the records relating to the same unit. There is a rich literature on record linkage which is beyond the scope of this paper \cite{fellegi1969theory,winkler1999state,herzog2010record,sayers2016probabilistic,deepak2018linking,komarova2018identification}. 
In the second scenario, the two samples are selected from the same population but have no common unit. In this case, a statistical matching framework has been proposed in survey studies, which finds a matched pair of units according to the shared variable $Z$, then imputes the missing value for one unit using the observed value from its matched counterpart  \cite{radner1980report,d2006statistical,ridder2007econometrics,d2015integration,yang2020statistical}. Validity of the statistical matching approach depends on the conditional independence assumption that conditional on the shared  variable $Z$, the potentially missing variables $Y$ and $X$ are independent. Under this assumption, 
matching on $Z$ is sufficient to impute $Y$ in $S=1$ sample regardless of whether $X$ are the same. A similar argument holds for imputation in the $S=0$ sample.

\subsubsection{Data combinition in regression analysis}
\cite{evans2018doubly} studied a different problem of estimating the regression coefficient of a correctly specified model $E[\yy\mid \zz,\xx;\theta]$ when both samples are i.i.d. random samples of the same population. Selection exchangeability and positivity were assumed similar to Section~\ref{sec:moment}, while no assumption on separable moments \cite{graham2016efficient,shu2020improved} or conditional independence \cite{ridder2007econometrics} introduced in previous sections was made.
In this setting, identification of $\theta$ can be hard even under linear models, which has been discussed in \cite{pacini2019two}, \cite{yang2020statistical}, and \cite{Miao2020commentary}. 
\cite{evans2018doubly} proposed a doubly robust estimator for $\theta$ that solves 
$\sum_{i=1}^{n} \widehat{U}(S_i,\yy_i,\xx_i,\zz_i;\theta)/n=0$ where 
\begin{equation}
	\begin{split}
		U(S,\yy,\xx,\zz;\theta)=
		g(\zz)\Big\{
		&\frac{S}{P(S=1\mid \zz)}\{\yy-E[\yy\mid \zz;\theta]\}+\\
		&\frac{1-S}{P(S=0\mid \zz)}\{E[\yy\mid \zz;\theta]-E[\yy\mid \zz,\xx;\theta]\}
		\Big\},
	\end{split}
\label{eq:evans}
\end{equation}
where $g(\cdot)$ is of the same dimension as $\theta$. The doubly robust estimator remains consistent under misspecification of either $f(\xx\mid \zz)$ or $P(S=1\mid \zz)$. Therefore,  an IPW estimator can be obtained by misspecifying $f(\xx\mid \zz)$ as zero, i.e., by substituting $E[\yy\mid \zz]$ with zero in Eq. (\ref{eq:evans}), while an imputation estimator can be obtained by substituting $P(S=1\mid \zz)$ with 0.5 in Eq. (\ref{eq:evans}).

\subsection{Two-sample instrumental variable and Mendelian randomization \label{sec:tsiv}}
An important setting  of data combination problem is the two-sample instrumental variable methods. 
An instrumental variable is an exogenous variable known to satisfy the following three core assumptions: (I) the IV must be associated with the treatment; (II) the IV must not have a direct effect on the outcome that is not mediated by the treatment; (III) the IV must be independent of unmeasured confounders. 
The IV approach is one of the most frequently used methods to mitigate unmeasured confounding denoted as $U$. 
It turns out that the causal effect can be estimated by combining information from two data sources. Let $Z$ denote an instrumental variable. The two-sample IV estimation concerns the scenario when $(Z, A, X, S=1)$ 
are available in one data source and $(Z, Y, X, S=0)$
are available in a separate data source, with $(Z,X)$ shared by the two datasets. No complete data on all variables $(Z, Y, A, X)$ are available.  In the following we will suppress the measured covariates $X$ to simplify notation, and all arguments are made implicitly conditional on $X$.  

We first consider the case of a binary treatment. 
Assuming that $U$ does not modify the causal effect of $A$ at the individual level, i.e., $Y=h(\epsilon) A+g(U,\epsilon)$, the ATE is identified by $ATE=E[h(\epsilon)]=cov(Z,Y)/cov(Z,A)$. 
Hence common IV methods often estimate the effect of the treatment  using  the  IV-outcome and IV-treatment  associations.
The numerator and denominator can be separately estimated from two distinct samples if both are random samples of the same target population. 
In a general case where $A$ is not necessarily binary and could be a vector, the most common IV approach assumes $Y=\beta A+\epsilon_Y$, and $A=\gamma Z+\epsilon_A$, and the IV estimator is given by $\widehat{\beta}=\widehat{\text{cov}}(Z,A)\inv \widehat{\text{cov}}(Z,Y)$, where $\widehat{\text{cov}}(\cdot,\cdot)$ denotes the sample covariance matrix. 
In the one-sample setting, the IV estimator is equivalent to a two-stage least squares (2SLS) estimator obtained by first regressing $A$ on $Z$, and then regressing $Y$ on $\widehat{A}$, the fitted values of $A$.
\cite{angrist1992effect} and \cite{arellano1992female} showed that the IV estimator  can be obtained by computing $\widehat{\text{cov}}(Z,A)$ based on the $S=1$ sample and  computing $\widehat{\text{cov}}(Z,Y)$ based on the $S=0$ sample, referred to as the two-sample IV estimator.
\cite{klevmarken1982missing} and \cite{angrist1995split} showed that the 2SLS can also be separately carried out using two samples, referred to as the two-sample two-stage least squares (TS2SLS) estimation \cite{bjorklund1997intergenerational}.  In the first stage $A$ is regressed on $Z$ using the $S=1$ sample, and the estimates are then combined with observations on $Z$ in the $S=0$ sample to form $\widehat{A}$. 
In the second stage, $Y$ is regressed on $\widehat{A}$. 
\cite{inoue2010two} pointed out that the equivalence of IV and 2SLS estimation in the one-sample setting does not hold in the two-sample setting.  In fact, TS2SLS is more efficient than two-sample IV because it implicitly corrects for differences in the distribution of $Z$ between the two samples.

The above classical two-sample IV methods often assume that the two samples are compatible with the same observed data distribution $f(Z,Y,A)$. 
However it is found that the common variable, i.e., the IV, can have different distributions between the two samples, i.e. $f(Z\mid S=1)\neq f(Z\mid S=0)$.  \cite{graham2016efficient} modeled the selection probability, $P(S=1\mid Z)$, parametrically and developed a doubly robust and locally efficient estimator which can be applied in more general data combination problems. Similar methods proposed in \cite{shu2020improved}, detailed in Section~\ref{sec:datafusion}, were also applied to the two-sample IV problem. 
\cite{sun2018semiparametric} established sufficient conditions for nonparametric identification of the ATE allowing for heterogeneous samples, derived the efficiency bound for estimating the ATE, and proposed a multiply robust and locally efficient estimator for estimation and inference.

Using genetic variants as IVs, two-sample Mendelian randomization (MR) methods have also been studied recently, which leverage publicly available summary statistics on genetic instrument-treatment and genetic instrument-outcome associations typically obtained from genome-wide association studies (GWAS) \cite{davey2003mendelian,pierce2013efficient,davey2014mendelian,lawlor2016commentary,zhu2018causal,davey2003mendelian,spiller2019software}. 
Although simple and convenient, the traditional two-sample MR methods typically rely on valid instruments. Methods robust to invalid instruments have been studied  \cite{bowden2015mendelian,bowden2016consistent,hartwig2017robust,li2017mendelian,zhao2020statistical,sanderson2020testing}, and extension to the setting of weak instruments has also been studied \cite{burgess2016bias,wang2019weak,sanderson2020testing}. 
\cite{zhao2019two} further considered the scenario when the sample compatibility assumption is violated and proposed methods that are robust to heterogeneous samples.

\subsection{Other causal inference problems\label{sec:datafusion_causal}}
\cite{fan2014identifying} studied the scenario when the shared variable is the treatment variable. More specifically, $(Y,A,X)$ are partially observed from two separate datasets: the outcome dataset contains $(A,Y,S=1)$, while the demographics dataset contains $(A,X,S=0)$. 
In this case, $E[Y\mid A,X]$ is not identified from neither dataset unless one is willing to make additional identification assumptions. 
Nevertheless, \cite{fan2014identifying} established sharp bounds for $E[Y(a)]$ via bounding its inverse probability weighting representation $E[\mathbbm{I}(A=a)Y/P(A=a\mid X)]$ under a continuous version of the classical monotone rearrangement inequality \cite{hardy1952inequalities,cambanis1976inequalities}. Other related works include \cite{manski2000identification}, \cite{cross2002regressions}, and \cite{ridder2007econometrics}.

A more general setting is studied in \cite{li2020causal} assuming $K+1$ datasets. Specifically, let $X=(X_{\scriptscriptstyle 1},X_{\scriptscriptstyle 2},\dots,X_K)$, $S\in\{1,\dots,K+1\}$ indicate each dataset, and $D_k$, $k=1,\dots,K+1$ denote the set of observed variable in the $k$-th dataset, with $D_{\scriptscriptstyle 1}=(A,Y,X_{\scriptscriptstyle 1},S=1),~D_{\scriptscriptstyle 2}=(A,Y,X_{\scriptscriptstyle 2},S=2),\dots,~D_K=(A,Y,X_K,S=K),~D_{K+1}=(X,S=K+1)$. Assuming that $Y(a)\ind A\mid X$, $S$ is randomly assigned, and $E[Y\mid A,X;\beta]$ is linear and additive, \cite{li2020causal} showed that the coefficient of $A$, which is the ATE under linear additive model, is identifiable by combining summary-level statistics obtained from the separate datasets.

\section{Other settings of data integration in causal inference\label{sec:others}}
\subsection{Distributed data setting}
Meta-analysis has a long history in integration of the results from multiple clinical trials with no access to individual-level trial data \cite{dersimonian1986meta,dersimonian2015meta}. Recently, another widely studied topic is the analysis of distributed data where individual-level observational data are not shareable due to privacy concerns \cite{toh2020analytic}. 
This is increasingly needed in multidatabase or multicenter study of comparative effectiveness and safety of medical products using  real-world  data such as electronic health records data. Each data partner can share a summary-level dataset with the analysis center. A few methods have been proposed and we summarize them ordered by the amount of information shared. The first method is to reduce the dimension of measured confounders using the propensity score or the prognostic score \cite{rosenbaum1983central,hansen2008prognostic}, then share individual-level treatment, outcome, and score with the analysis center to apply propensity score methods \cite{rassen2012using,shi2020safety}. The second method is to aggregate subjects into cells defined by confounders or the propensity score strata, then adjust for confounding based on counts of subjects in each cell \cite{cook1989performance,rassen2010privacy,shu2020inverse}. Propensity score matching within each data partner can be done prior to the aggregation \cite{toh2013confounding,yoshida2018comparison}. The third one is distributed regression \cite{zhang2013divide,toh2018combining}, and the last one is meta-analysis of site-specific results \cite{toh2013confounding}. 

\subsection{Bayesian causal inference}
Bayesian framework can naturally facilitate the borrowing of prior information across data sources  \cite{ibrahim2000power,gelman2006prior,hobbs2011hierarchical,kaizer2018bayesian}. 
\cite{boatman2020borrowing}  studied the problem of estimating causal effects from a primary source and borrowing from any number of supplemental sources when data on outcome, treatment, and confounders are available in all data sources. When some confounders are unmeasured in a large main dataset but are available in  a small validation dataset, a missing data perspective have been used to impute the missing covariates \cite{gelman1998not,jackson2009bayesian,murray2016multiple}. 
When the number of missing covariates in the main study is large relative to the sample size of the validation study, \cite{antonelli2017guided} proposed a Bayesian approach to estimate the ATE in the main study that combines Bayesian variable selection and missing data imputation, allowing for heterogeneous treatment effects between the main and validation studies.
\cite{comment2019bayesian} proposed to use informative priors on quantities related to the unmeasured confounding bias in a range of settings including both static and dynamic treatment regimes as well as  treatment-induced mediator-outcome confounding.

\subsection{Causal discovery}
Data integration has also been studied in causal discovery, which aims to learn the causal relations between variables of a system, using multiple heterogeneous datasets  that measure the system 
under different environments or experimental conditions and with different sets of variables.
There are two main types of methods. The first type pools data from different experiments to learn a context-independent causal graph of the system \cite{cooper1999causal,tian2001causal,eaton2007exact,peters2016causal,zhang2017causal}. For example,  \cite{peters2016causal} provided an invariant prediction method built on the idea that the conditional distribution of the outcome given the direct causes is invariant across different experimental conditions. \cite{mooij2020joint} proposed to take into account context variables that discriminate the different datasets in standard causal discovery methods applied to the pooled data. The second type derives statistics or constraints from each context separately without pooling data and combines them to learn a single graph \cite{claassen2010causal,tillman2011learning,triantafillou2015constraint}. 

\section{Discussion\label{sec:discussion}}
In this paper, we reviewed a collection of data integration methods in causal inference.
A common perspective views data integration in causal inference as a missing data problem where the study sample is a subset of the target population. This problem is referred to as generalizability or verify-in-sample. We summarize the data missing patterns in Sections~\ref{sec:trial}-\ref{sec:nocompleteddata} in Table~\ref{tbl:msng}. 
Another setting increasingly recognized is when the study sample and the target population are partially- or non-overlapping, in which selection exchangeability requires that the variables that determine study inclusion/exclusion should not be predictive of the outcome or at least does not modify the treatment effect. This problem is referred to as transportability or verify-out-of-sample \cite{chen2008semiparametric,colnet2020causal,dahabreh2020extending,degtiar2021review}. 
We summarized causal inference methods  under both scenarios and their applications in important real-world problems including combining clinical trial with external information, correcting for unmeasured confounding in observational study using auxiliary or trial data, two-sample Mendelian randomization, and distributed data network. Majority of the methods relies on some form of exchangeability/homogeneity across different data sources, hence sensitivity to violation of exchangeability assumptions should be routinely conducted. In addition, identification strategies in complex settings such as when no single sample contains all relevant variables have not been fully explored, and connection to the covariate shift problem in machine learning has yet to be fully studied.

\begin{table}[!h]
		\resizebox{\textwidth}{!}{
	\begin{tabular}{|c|c|c|c|c|c|c|c|c|c|c|c|c|c|c|c|c|c|c|c|}
		\hline 
		\bf Section & \multicolumn{3}{c|}{\bf \ref{sec:general}} & \multicolumn{4}{c|}{\bf \ref{sec:correctobs}} & \multicolumn{3}{c|}{\bf \ref{sec:external}} & \multicolumn{3}{c|}{\bf \ref{sec:datafusion}} & \multicolumn{3}{c|}{\bf \ref{sec:tsiv}} & \multicolumn{3}{c|}{\bf \ref{sec:datafusion_causal}}\tabularnewline
		\hline 
		\bf Variable & $Y$ & $A$ & $X$ & $Y$ & $A$ & $X_{1}$ & $U$ & $Y$ & $A$ & $X$ & $Y$ & $Z$ & $X$ & $Y$ & $Z$ & $A$ & $Y$ & $A$ & $X$\tabularnewline
		\hline 
		\bf $S=1$ & \Checkmark{} & \Checkmark{} & \Checkmark{} & \Checkmark{} & \Checkmark{} & \Checkmark{} & \Checkmark{}{\large/}\XSolidBrush{} & \Checkmark{} & 1 & \Checkmark{} & \Checkmark{} & \Checkmark{} & \XSolidBrush{} & \Checkmark{} & \Checkmark{} & \XSolidBrush{} & \Checkmark{} & \Checkmark{} & \XSolidBrush{}\tabularnewline
		\hline 
		\bf $S=0$ & \XSolidBrush{} & \XSolidBrush{} & \Checkmark{} & \Checkmark{} & \Checkmark{} & \Checkmark{} & \XSolidBrush{} & \Checkmark{} & 0 & \Checkmark{} & \XSolidBrush{} & \Checkmark{} & \Checkmark{} & \XSolidBrush{} & \Checkmark{} & \Checkmark{} & \XSolidBrush{} & \Checkmark{} & \Checkmark{}\tabularnewline
		\hline 
	\end{tabular}}
	\caption{\label{tbl:msng} Data missing patterns in the major settings discussed in Sections~\ref{sec:trial}-\ref{sec:nocompleteddata}. Here \Checkmark{} stands for observed and \XSolidBrush{} stands for unobserved, and \Checkmark{}{\large/}\XSolidBrush{} indicates different settings considered by different papers.}
\end{table}
\clearpage
\bibliographystyle{unsrt}
\bibliography{sample}
\end{spacing}
\end{document}